\documentstyle[12pt,aasms4]{article}
\def\lapprox{\hbox{\lower .8ex\hbox{$\,\buildrel < \over\sim\,$}}}
\def\gapprox{\hbox{\lower .8ex\hbox{$\,\buildrel > \over\sim\,$}}}

\begin{document}

\title
{Identification of the companion stars of Type Ia supernovae}

\author
{R. Canal\altaffilmark{1}, J. M\'endez\altaffilmark{1,2}, and 
P. Ruiz--Lapuente\altaffilmark{1,3}}

\altaffiltext
{1}{Department of Astronomy, University of Barcelona, Mart\'\i\ i Franqu\'es
1, E--08028 Barcelona, Spain. E--mail: pilar@am.ub.es, ramon@am.ub.es}

\altaffiltext
{2}{Isaac Newton Group, Apartado 321, Santa Cruz de La Palma, E--38780 
Tenerife,Spain. E--mail: jmg@ing.iac.es}

\altaffiltext
{3}{Max--Planck--Institut f\"ur Astrophysik, Karl--Schwarzschild--Strasse 1,
D--85740 Garching, Germany. E--mail: pilar@MPA--Garching.MPG.DE}

\slugcomment{{\it Running title:} Type Ia supernova companions}

\begin{abstract}
The nature of the binary systems giving rise to Type Ia supernovae (SNeIa) 
remains an unsolved problem. In this {\it Letter} we calculate, from the 
statistics of initial conditions (masses and binary separations), the mass, 
luminosity, and velocity distributions of the possible binary companions 
(main--sequence star, subgiant, red giant) following the explosion of the 
white dwarf which gives rise to the SNeIa. Those companions could be 
detected from either their proper or their radial motions, by means of 
high--precision astrometric and radial--velocity measurements in young, 
nearby supernova remnants. Peculiar velocities typically ranging from 100 to 
450 km s$^{-1}$ should be expected, which places proper--motion measurements
within reach of HST instruments and makes radial--velocity ones feasible with
2.5--4m class telescopes from the ground. Detections would solve the 
long--standing problem of which kind of binaries do produce SNeIa and clear 
up the way to accurate physical modeling of the explosions. 
\end{abstract}


\section{Introduction}
Type Ia supernovae (SNeIa) are unanimously attributed, nowadays, to the 
thermonuclear explosion of a white dwarf (WD) due to accretion of matter 
from a companion in a close binary system. The mass--accreting WD is also 
generally thought to be made of carbon and oxygen (CO WD). There the agreement
stops, and different kinds of systems are proposed as candidates to SNeIa 
progenitors (see Iben 1997, for a review).

The nature of the WD companion, the mass--donor, is unclear:
it could either be another WD (double--degenerate or DD systems) or a still 
thermonuclearly evolving star (single--degenerate or SD systems). The 
evolutionary stage of the companion in the SD systems could be anything from a
main--sequence star to a supergiant,  and mass accretion proceed via 
Roche--lobe overflow (driven by thermonuclear evolution of the 
companion, magnetic braking, or gravitational wave radiation) or by capture 
of matter from stellar wind. All that also bears on the explosion 
mechanism: on how the explosion is triggered and how it develops. Those 
uncertainties still raise doubts as to the use of SNeIa as {\it calibrated 
candles} to probe the dynamics of the Universe, the calibration being 
purely empirical and based on the nearby SNeIa sample (see Drell, Loredo, 
\& Wasserman 2000, for instance).

Recently, evidence has been gathered against the merging of WDs giving rise
to SNeIa, from the high efficiencies found in producing SNeIa out of star 
formation. These efficiencies have now been measured at several
redshifts and appear to be similar, every $\sim 700\ M_{\odot}$ going into
star formation giving rise to 1 SNeIa (Ruiz--Lapuente \& Canal 2000).

No previously observed system has ever been identified as the progenitor of 
any SNeIa. The historical SNeIa took place at a time when 
astronomical instrumentation was rudimentary. They can still give us, 
nonetheless, important and maybe decisive clues on the nature of the 
long--sought SNeIa progenitor systems (Ruiz--Lapuente 1997). If they were DD 
systems, there should be no companion left since it is destroyed in the 
accretion process already. On the contrary, in SD systems the companion can 
survive the explosion in most cases and show characteristics allowing its 
identification.

Of the historical SN, from the records on lights curves only two have
been identified as SNeIa: SN 1006 and Tycho's SN (SN 1572). From their X--ray 
morphology (spherically--symmetric shell), X--ray spectra (showing high 
abundances of Fe, Ni, Si, Ca, Ar), Fe enrichment in their optical filaments, 
and lack of any central X--ray source, a number of nearby, young Galactic 
supernova remnants (SNR) can be attributed to SNeIa 
(Ruiz--Lapuente et al. 2000). As we will see, precise enough astrometric and 
radial--velocity measurements of the central regions of such SNR should allow 
detection of the SNeIa companions from either their large proper motions or 
high radial velocities, due to disruption of the binary orbit plus kick from 
the impact of the SN ejecta on them. For SNR ages $\lapprox 10^{4}\ yr$, the 
companions should still exhibit in their outer layers the consequences of the 
strong perturbation of thermal equilibrium from partial stripping of the 
envelopes and deposition of energy from the ejecta into the layers that 
remain bound, and they may also show chemical contamination from the SNeIa 
ejecta.

\section{Modeling, Results, and Discussion}

The SD scenarios thus far proposed to explain the origin of SNeIa involve 
different kinds of systems. We will only consider Ch--mass explosions, which
are generally thought to produce the bulk of SNeIa, so the SN ejecta 
impinging on the companion will always be identical (differences due to the 
explosion mechanism should be minor and will be left aside). Depending on the 
kind of system, however, the mass and evolutionary stage of the companion, as 
well as the separation between the two stars at the time of explosion, will 
be different and that will determine the space velocity and the mass, 
luminosity, surface temperature, and surface abundances of the surviving star.

We will consider the cases of a main--sequence companion filling its 
Roche lobe due to angular--momentum loss via magnetic braking (cataclysmic 
variable or CV system), and that of a subgiant or red-giant  
companion filling its Roche lobe due to thermonuclear evolution 
(cataclysmic--like or CLS system). The latter are thought to show up as 
luminous supersoft X--ray sources before exploding as SNeIa. In the modeling 
of the CLS systems, the stabilizing effects on mass transfer of the strong 
wind generated in the WD surface have been taken into account as in Hachisu, 
Kato, and Nomoto (1996). The cosmic SNeIa rates thus predicted give the best 
agreement with the observations (Ruiz--Lapuente \& Canal 2000).

Our purpose here is to derive the statistical distribution of velocities, 
masses, and luminosities of the companions after explosion.  
We first run and updated version of the Monte Carlo scenario code we used to 
model SNeIa rates in previous work (Ruiz--Lapuente, Burkert, \& Canal 1995;
Canal, Ruiz--Lapuente, \& Burkert 1996; Ruiz--Lapuente \& Canal 1998), to 
obtain the distributions of companion masses and orbital separations at the 
time of explosion, for the systems considered. Those are displayed in Figure 1
for the CV systems (main--sequence, MS companion) and the CLS systems 
(distinguishing between subgiant SG, and red--giant RG companions). Note, 
in the upper panel, that the mass distributions of the SG and RG stars in 
the CLS scenario are the same, the SG case corresponding to the narrow 
range of final separations shown in the lower panel whilst the RG case
covers a much broader range, with an almost flat distribution (same panel). 
In the CV scenario, the much narrower distribution of final masses, peaking
at $\sim 0.6 M_{\odot}$, means an also narrow distribution of final radii, 
the companion being still in the MS. Making contact with the Roche lobe
while having a companion (the WD) that has always the Ch--mass produces
the sharp peak in the distribution of final separations that we see in the 
lower panel. 
  
From the initial conditions above, the effects of the impact on the companion 
are calculated in a semianalytical approximation equivalent to that used 
by Wheeler, Lecar, \& McKee (1975), based on previous work by Colgate (1970). 
The companion star is divided into concentric cylinders and the ejecta are 
treated as a plane slab. If the momemtum incident on a cylinder can
accelerate it to the scape velocity, the corresponding mass is directly
stripped. A strong shock wave is driven into the rest of the star: that 
will impart momentum to it (kick) and the heating will make the internal 
energy to exceed the gravitational potential energy in extra layers which 
will thus be expelled. The approximation is made that the fluid velocity
just behind the shock is equal to the mean velocity of all the shocked 
stellar material. The energy per unit mass rapidly increases as the mass per
unit area in the cylindrical shells decreases: energy deposition is thus
concentrated in the outer layers, which ensures that the companion is never
entirely disrupted. The results have been compared with the two--dimensional 
hydrodynamic simulations of Marietta, Burrows, \& Fryxell (2000). 
Especifically, this has been done for the cases of a $1\ M_{\odot}$ MS 
companion at separation $a = 3\ R_{\odot}$, a $1.1\ M_{\odot}$ SG at $a = 
4.9\ R_{\odot}$, a $2.1\ M_{\odot}$ SG at $a = 6.5\ R_{\odot}$, and a 
$1\ M_{\odot}$ RG at $a = 430\ R_{\odot}$. The agrement (maximum discrepancy 
below 20\%) is good enough to confirm our approximation as a reliable tool 
for the statistical evaluation of the characteristics of the SNeIa companions 
after explosion, which would be hard to do by means of detailed hydrodynamic 
calculations. Note that the combinations of companion masses and 
separations at the time of explosion in Marietta, Burrows, \& Fryxell (2000) 
(see their Table 4) are just adapted from examples proposed by Livio \& Truran 
(1992) (CV system) and Li \& van den Heuvel (1997) (CLS system), whereas in 
the present work the actual distributions of those parameters have been 
calculated for each scenario. That is why the cases above (with the
exception of the $1.1\ M_{\odot}$ SG companion at $a = 4.9\ R_{\odot}$) are
not typical of the predicted distributions. However, Marietta, Burrows, \& 
Fryxell (2000) have compared they results with an aproximation similar to 
ours and find best agreement for binaries near Roche lobe overflow, which is 
always our case. That reinforces the significance of our comparisons.  

The structures of the MS and SG companions prior to explosion have been 
obtained by homologous transformations of a $1 M_{\odot}$ model in the 
corresponding stages. Typically, 10--20 \% of the companion mass is lost. The 
RG companions are modelled as a compact He core surrounded by an extended 
H--rich envelope, the latter having the structure of a $n = 1.5$ polytrope. 
The relationship between RG mass, radius at the time of explosion, and core 
mass, has been taken from the stellar evolution calculations by Politano 
(1988), which cover the initial mass range $0.1\ M_{\odot}\leq M\leq 10\ 
M_{\odot}$. We find that in all cases the RG envelope is completely stripped 
off, as in the 2D simulations of Livne, Tuchman, \& Wheeler (1992). Marietta, 
Burrows, \& Fryxell (2000), in their recent 2D hydrodynamic modeling find, 
however, that a small fraction of the envelope is retained, which would 
greatly help in the identification of possible SNeIa companions when 
exploring the central regions of young, nearby SNR, as discussed below.

The kick velocities imparted to the companions by collision wih the SNeIa 
debris are also calculated, but they are always much smaller than the
typical orbital velocities at the time of explosion, for the three types of
companions (MS, SG, RG). The predicted distributions of total velocities 
(orbital plus kick) are shown in Figure 2.

The results corresponding to the most typical combinations of mass and 
orbital separation, for each type of companion, are shown in Table 1. In
column 2 we give the apparent visual magnitude $m_{V}$ expected (for 
distances of 1 kpc and 5 kpc). In column 4, the average radial velocities
$v_{r}$ (equal to the average transversal velocities $v_{t}$, since the 
total velocities should be distributed at random with respect to the observer's
line of sight). In column 2 is the proper motion $\mu$ (in arcsec 
yr$^{-1}$), and in column 5 the maximum angular distance $\theta$ (in arcmin) 
from the explosion site that would be reached $10^{3}\ yr$ after the SN 
event. Given in the footnotes to Table 1 are the pre--explosion and 
post--explosion masses for the three cases. We must stress here that the 
apparent visual magnitudes given in Table 1 are calculated from the 
luminosities that the companions would reach {\it after relaxing to thermal 
equilibrium}, and thus are {\it lower limits} to the actual brightnesses to 
be expected up to a few thousand years after explosion. It is clear that the 
strong shock wave will not only unbind a fraction of the outer layers and 
give a global kick to the companion but it will also strongly perturb the 
hydrostatic and thermal equilibrium of the fraction that remains bound. 
Mechanical equilibrium should be restored on a hydrodynamic time scale and 
thermal equilibrium on a longer, Kelvin--Helmholtz time scale. Indeed, 
Marietta, Burrows, \& Fryxell (2000) find in their 2D numerical simulations 
that after impact the MS and SG companions would swell up, their luminosities 
rapidly increasing by factors of up to a few thousand, to relax back to 
luminosities of the order shown in Table 1 in $\sim 10^{4}\ yr$. The RG 
companions would retain a residual H--rich envelope, still able to feed the 
H--burning shell and would evolve away from the RG branch at constant 
luminosity and increasing effective temperature, on a time scale of 
$\sim 10^{4} - 10^{5}\ yr$. Therefore, SNeIa companions left by recent, 
nearby explosions, should in fact be detectable at magnitudes significantly 
brighter than those shown in Table 1. To illustrate this we also give, 
within parentheses, the magnitudes expected for the MS companions (the less
luminous ones) at $t\sim 10^{3}\ yr$ after explosion. The peculiar velocities 
predicted are one order of magnitude larger that the systemic velocities of 
field stars at the corresponding distances. The angular sizes of the regions 
to be explored (given by $\theta$ in Table 1) ensure that the number of stars 
bright enough to be candidates to SNeIa companions, contained within their 
boundaries, is quite small. Detection of proper motion at the level of 
0.01 -- 0.005 arcsec yr$^{-1}$, even at the magnitudes estimated in Table 1, 
is within the reach of the instruments FGS1R(brighter targets) and WFPC2 
(fainter targets) on board the {\it Hubble Space Telescope} (Ruiz--Lapuente et 
al. 2000). Failure to detect any candidates would be strong evidence that 
none is left after the SNeIa explosions and almost completely discard the SD 
system hypothesis.

The possible candidates selected from their high proper (or radial) motion 
should then be spectroscopically analyzed. As we have seen, MS and SG 
candidates should be swollen and overluminous and RG candidates would have 
high effective temperatures. Moreover, Marietta, Burrows, \& Fryxell (2000) 
find that whereas no material from the ejecta should be accreted during 
impact, there is a possibility of later accretion of material enriched in Fe 
and intermediate--mass elements from the low--velocity tail of the ejecta, 
long after impact.

There are still many uncertainties concerning close binary evolution, binary
mass transfer, and the process of mass growth of a WD from accretion until
it reaches the Ch--mass. To which extent could those uncertainties affect our
result that any SNeIa companion should be moving at high velocities after
the explosion and be luminous enough to be detected in the central regions 
of young, nearby SNR?      

In the case of the CV systems, the companions are MS stars that continuosly 
fill their Roche lobes due to angular momentum loss by magnetic braking (plus 
emission of gravitational waves when the orbits become very close). The WD
mass being equal to the Ch--mass, separation is only a function of companion
mass, which in turn is just the initial mass of the star minus the fraction
transferred to the WD and the (minor) fraction lost to stellar wind. The
efficiency of magnetic braking is here the main unknown, but it hardly affects
the distribution of final masses (and thus separations) shown in Figure 1 nor
the values predicted in Table 1.

Although the CLS systems involve companions in a more advanced evolutionary
stage than the CV systems, our main conclusions are equally robust. When the
companion is a SG, the case is as clear--cut as for the CV systems. For the
RG companions, the only worrying possibility, that the companion velocities 
actually were much lower than in our modeling, can be reliably excluded: 
since $v_{orb}\propto a_{f}^{-1/2} (M_{1} + M_{2})^{1/2}$ and the WD mass is 
always equal to the Ch-mass, we approximately have $v_{orb}\propto 
a_{f}^{-1/2} (M_{1} + 1.4)^{1/2}$, $M_{1}$ being the companion's mass. 
Increasing $a_{f}$ (and, less efficiently, lowering $M_{1}$) would thus 
decrease $v_{orb}$. But the companion must be filling its Roche lobe $R_{L}$, 
and we also have (Paczy\'nski 1971) that $R_{L}\propto a_{f}\ (0.351 + 
log\ M_{1})$ (where in the numerical constant we have again taken 
$M_{2}\simeq M_{Ch}$). Increasing  $a_{f}$ should, therefore, increase 
$R_{L}$ faster than it would decrease $v_{orb}$ (decreasing  $M_{1}$ cannot 
go very far, since the companion's initial  mass must be $M_{1}\gapprox 0.8\ 
M_{\odot}$ in order to have evolved from the MS in less than a Hubble time), 
and substantial increase postpones Roche--lobe overflow to the red supergiant 
(RSG) stage, when a common envelope would form and be ejected, that 
suppressing mass transfer to the WD altogether and leaving a DD 
system, no longer a SD one. 

A class of SD system that we have not considered here is the one in which
the WD accretes H--rich material from the wind of a RG or RSG that is not 
filling its Roche lobe, a symbiotic system (SS). There the orbital separation
could be larger and the orbital velocities smaller. However, the difficulty 
in making the WD grow to the Ch--mass through such a low--efficiency 
accretion process, avoiding explosive ignition of either H or He in the outer
layers (which requires high accretion rates), has discarded SS systems as
possible progenitors of the bulk of the SNeIa at least (Kenyon et al. 1993).

Finally, that the RG companions do lose their envelopes at most, and that
the SG and MS companions typically just lose a minor fraction of their masses
upon being hit by the SNeIa ejecta, is consistently found in all hydrodynamic 
simulations and semianalytical approximations. That upon impact and stripping
of the outer layers the new surface layers of the companions will be out
of thermal equilibrium, and that the objects should thus initially appear 
overluminous as compared with the equilibrium state they will approach on a 
thermal time scale, seems hardly disputable as well.          

\section{Conclusions}

We have added to the results from existing hydrodynamical studies of the 
impact of SNeIa into their companion stars the statistical evaluation of the 
distribution of velocities to be expected for different kinds of 
companions. We have shown that they should be moving at velocities much 
higher than those of the surrounding stars: for RG companions there is a
probability $P > 90\%$ that the velocities are higher than $100\ km\ s^{-1}$; 
for SG $P > 99\%$ that $v > 200\ km\ s^{-1}$, and for MS $P > 99\%$ that
$v > 450\ km\ s^{-1}$. Such velocities make their identification possible in 
the central regions of young, nearby SNR of SNeIa origin. 
For distances $d\lapprox 5 - 8\ kpc$ and SNR ages $\lapprox 10^{3} - 
10^{4}\ yr$, detection of proper motion is within the reach of instruments 
on board the HST, and that of radial motions can be made with 2.5--4m class 
telescopes from the ground. The angular sizes of the regions to be surveyed 
are small enough that the numbers of stars within them, bright enough to be 
candidates to SNeIa companions, remains low. There are 8 Galactic SNR 
(including those of SN 1006 and SN 1572) that most likely formed from SNeIa 
explosions and whose distances and ages allow thorough searches 
(Ruiz--Lapuente et al. 2000).

Candidates selected from their peculiar velocities and positions close to the
centers of the SNR could be confirmed by spectroscopic characteristics 
arising from their being out of thermal equilibrium due to heating by the 
impact of the SN debris, according to recent hydrodynamic simulations 
(Marietta, Burrows, \& Fryxell 2000). Contamination by material accreted from 
the low--velocity tail of the ejecta is also possible.

Our conclusions are largely independent from the unknowns still affecting
close binary evolution, mass transfer, and growth of WDs up to the Ch--mass.

Therefore, it now appears feasible to test the SD system hypothesis for the
origin of SNeIa by means of observations of Galactic SNR. Detection of 
confirmed companions would not only validate the SD system hypothesis, but it 
would also allow physical modeling of SNeIa on a firmer basis and, through 
that, make much more reliable than at present the use of SNeIa in Cosmology.

\clearpage

\begin{table*}[htb]
\caption{Typical apparent magnitudes, proper motions, radial velocities, 
and maximum angular distance from explosion site (after 10$^{3}$ yr) of
SNeIa companions}
\label{table:1}
\newcommand{\m}{\hphantom{$-$}}
\newcommand{\cc}[1]{\multicolumn{1}{c}{#1}}
\renewcommand{\tabcolsep}{2pc} 
\begin{tabular}{@{}lllll}
\hline
Companion type & m$_{V}$ & $\mu$ (arcsec yr$^{-1}$) & v$_{r}$ (km s$^{-1}$) &
$\theta$ (arcmin) \\

and distance   &         &                          &              &     \\
\hline

               &         &                          &              &     \\
Main sequence $^{a}$  &  &                          &              &     \\
1 kpc          &  18.7 (11.2)$^{*}$  & 0.067        & 320          & 1.6 \\
5 kpc          &  22.2 (14.7)$^{*}$  & 0.013        & 320          & 0.3 \\

               &         &                          &              &     \\
Subgiant $^{b}$       &  &                          &              &     \\
1 kpc          &  12.6   & 0.038                    & 180          & 0.9 \\
5 kpc          &  16.1   & 0.008                    & 180          & 0.2 \\

               &         &                          &              &     \\
Red giant $^{c}$      &  &                          &              &     \\
1 kpc          &  10.5   & 0.015                    &  70          & 0.4 \\
5 kpc          &  14.0   & 0.003                    &  70          & 0.1 \\

               &         &                          &              &     \\  
\hline
\end{tabular}\\[2pt]
$^{a}$ Pre--explosion mass: $0.60\ M_{\odot}$; post--explosion mass: $0.51\  
M_{\odot}$.
$^{b}$ Pre--explosion mass: $1.00\ M_{\odot}$; post--explosion mass: $0.85\  
M_{\odot}$.
$^{c}$ Pre--explosion mass: $1.00\ M_{\odot}$; post--explosion mass: $0.33\  
M_{\odot}$ (electron--degenerate He core).

$^{*}$ Expected at $t\sim 10^{3}\ yr$ after explosion.
\end{table*}

\clearpage

\begin{figure}[hbtp]
\centerline{\epsfysize15cm\epsfbox{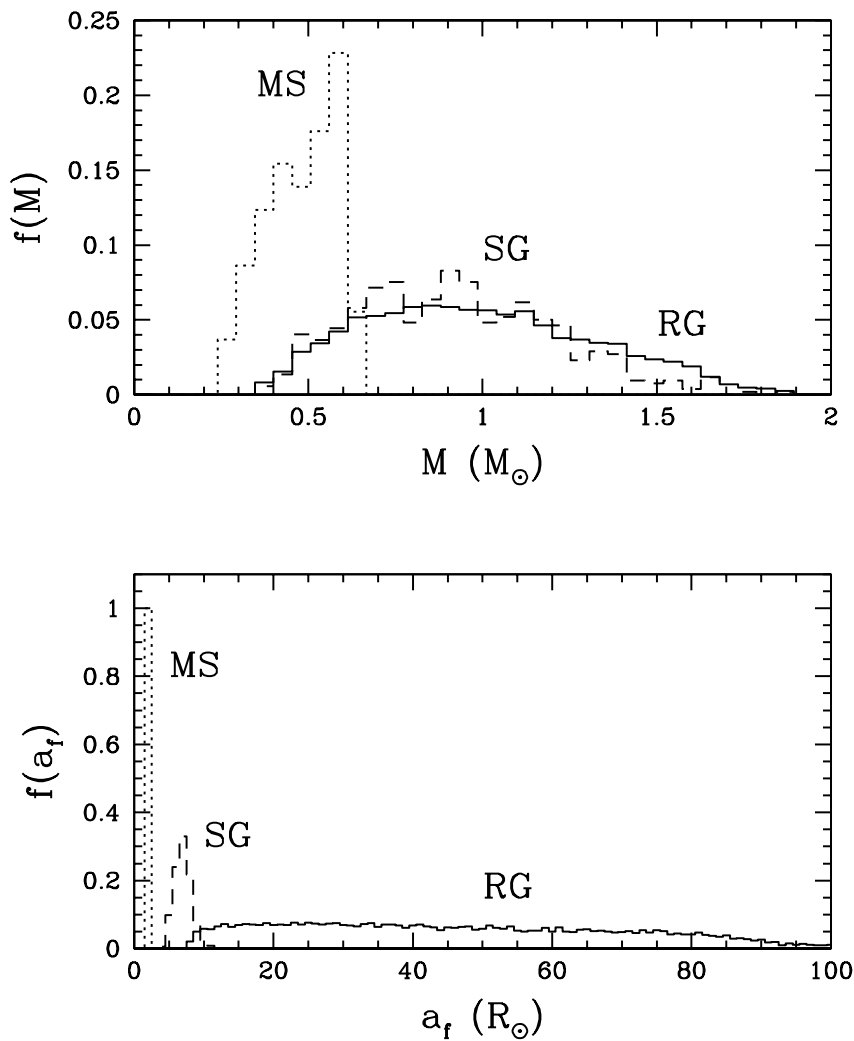}}
\nopagebreak[4]
\figcaption{Predicted distributions of SNeIa companion masses (upper panel) 
and of orbital separations (lower panel), just prior to SNeIa explosion, for
the two types of single--degenerate systems considered: a WD in a 
cataclysmic variable system, where the companion is a MS star (dotted line), 
and in a cataclysmic--like system, where the companion can be either a SG 
(dashed line) or a RG star (solid line). Note that in the lower panel the
values corresponding to the RG case have been multiplied by a factor 5, to
improve legibility.   
\label{fig1}}
\end{figure}

\begin{figure}[hbtp]
\centerline{\epsfysize15cm\epsfbox{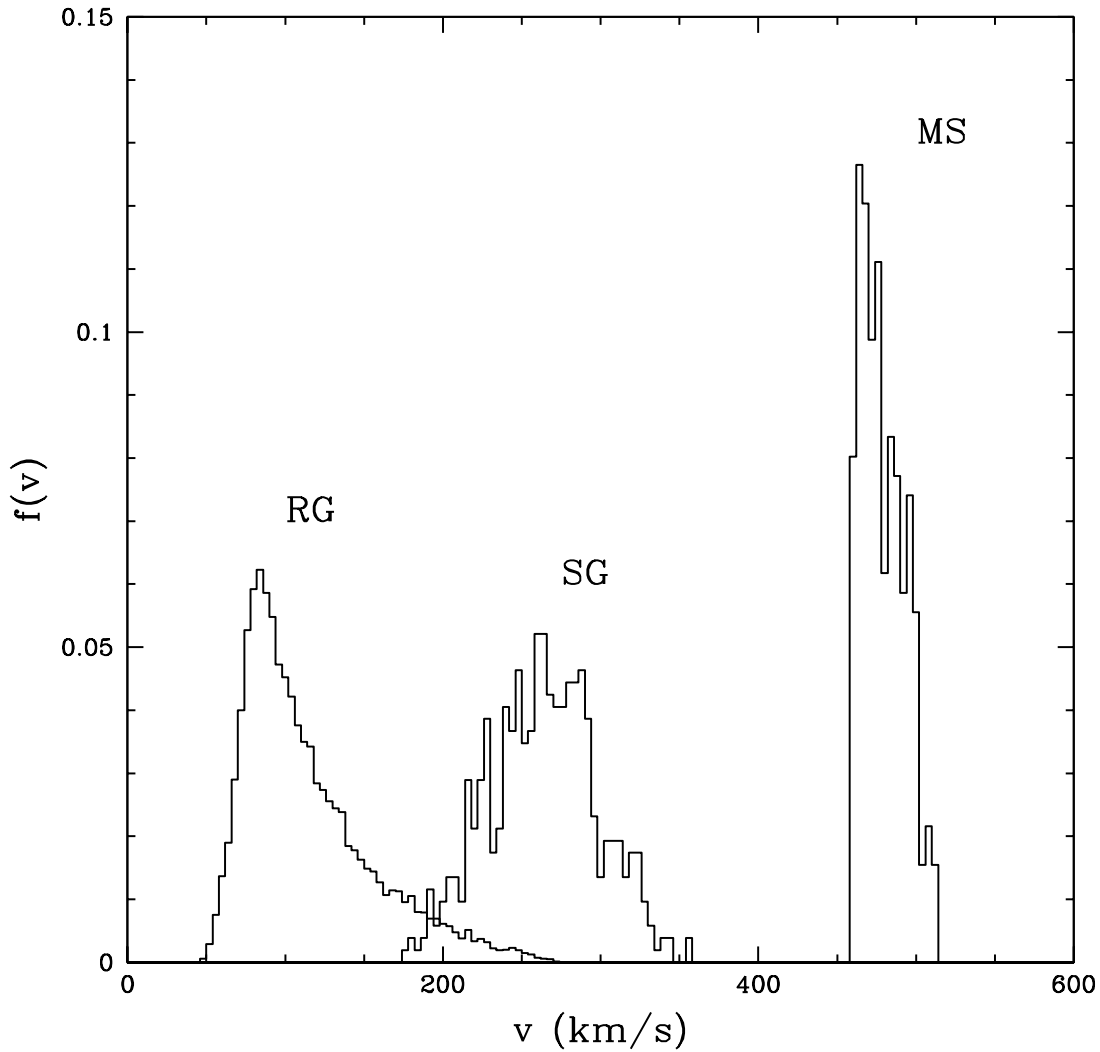}}
\nopagebreak[4]
\figcaption{Predicted velocity distributions of possible binary companions of
the exploding WD, following the SNeIa explosion, for the same 
single--degenerate systems as in Figure 1. Our statistical evaluation suggests
that if the companion star is a RG there is a probability $P > 90\%$ that it 
will move at a velocity larger than 100 km s$^{-1}$. For the SG, there is a 
$P > 99\%$ that it will move at a velocity larger than 200 km s$^{-1}$, and 
for the MS companion there is probability $P > 99\%$ that it will be moving 
at a velocity larger than 450 km s$^{-1}$. The chances that the central 
regions of young SNR produced by SNeIa contain a star moving at a velocity 
larger than 100 km s$^{-1}$ are thus very high if the SD scenario is true.
\label{fig2}}
\end{figure}

\end{document}